\documentclass[twocolumn,amsmath,amssymb,prl,aps]{revtex4}

\usepackage{graphicx}
\usepackage{color}

\newcommand{\nc}{\newcommand}
\nc{\ie}{\textit{i.e.} }
\nc{\del}{\partial}
\nc{\Cin}{C_{\rm{in}}}
\nc{\Cout}{C_{\rm{out}}}
\nc{\Epos}{E_{e^+}}
\nc{\fann}{f_{\mathrm{ann}}}
\nc{\ud}{\mathrm{d}}
\nc{\JM}{\bf JORDI:}

\begin{document}

\title{The positron density in the intergalactic medium and the galactic
 511 keV line}

\author{A.~Vecchio$^{1,2}$, A.C.~Vincent$^{1}$, J.~Miralda-Escud\'e$^{3,4}$ and C. Pe\~na-Garay$^{1}$\\
 \it{ $^1$ Instituto de Fisica Corpuscular, CSIC-UVEG, Valencia 46071 Spain\\ 
 $^2$Dipartimento di Fisica, Universit\`a della Calabria, 87036 Rende (CS), Italy \\
 $^3$ ICREA, Barcelona, Catalonia}\\
 $^4$ Institut de Ci\`encies del Cosmos, Universitat de Barcelona/IEEC, Barcelona 08028\\
}

\begin{abstract}
\noindent The 511 keV electron-positron annihilation line, most recently
characterized by the INTEGRAL/SPI experiment, is highly concentrated towards
 the Galactic centre. Its origin remains unknown despite decades of
scrutiny. We propose a novel scenario in which known extragalactic
positron sources such as radio jets of active galactic nuclei (AGN) fill
the intergalactic medium with MeV $e^+e^-$ pairs, which are then
accreted into the Milky Way. 
We show that interpreting the diffuse cosmic radio background (CRB) as arising
from radio sources with  characteristics similar to the observed cores
and radio lobes in powerful AGN jets suggests that the intergalactic
positron-to-electron ratio could be as high as $\sim 10^{-5}$, although this
can be decreased if the CRB is not all produced by pairs and if not all
positrons escape to the intergalactic medium.
Assuming an accretion rate of one solar mass per year of matter into the
Milky Way, a positron-to-electron ratio of only $\sim 10^{-6}$ is
already enough to account for much of the 511 keV emission of the
Galaxy. A simple spherical accretion
model predicts an emission profile highly peaked in the central bulge,
consistent with INTEGRAL observations. However, a realistic
model of accretion with angular momentum would likely imply a more
extended emission over the disk, with uncertainties depending on the
magnetic field structure and turbulence in the galactic halo.
\end{abstract}

\pacs{Valid PACS appear here} \maketitle

\maketitle

The long-standing problem of the origin of the Galactic 511 keV line
remains unsolved, four decades after its first detection.  
Observations by the INTEGRAL/SPI experiment imply the annihilation of
at least $\sim 2\times10^{43}$ positrons per second, mainly from a
spherical bulge at the galactic centre subtending an angle
of $\sim 10^\circ$ in the sky
\cite{Knodlseder:2005yq,Weidenspointner2008}. The bulge-to-disk ratio
of luminosities from $e^+e^-$ annihilation is $\sim$ 1.4
\cite{integral} and the ratio of the 511 keV line to
the lower energy continuum is consistent with $100\%$
annihilation via positronium formation \cite{Jean:2005af}.

Positrons are copiously produced in astrophysics. Stellar
nucleosynthesis and supernova explosions produce radioactive nuclides
that decay via $\beta^{+}$ emission, resulting in MeV--scale positrons. 
High-energy particles, either photons or cosmic rays, can produce
$\gg$ MeV positrons via pair production or pion decay. 
The rate of positron production in the Milky Way has been estimated for
a number of candidate sources, but none reproduce the correct morphology
(see \cite{review} for a comprehensive review).
Positrons produced by these sources
can travel long distances before annihilating, reducing the correlation
between the distribution of sources and the detection pattern, although
this does not lead to any concentration of the signal in the bulge
\cite{refId0}. In light of the difficulties to explain the properties of
the Galactic 511 keV emission with astrophysical sources, annihilation
or decay of dark matter has been invoked as an alternative source of
positrons and as an explanation of the large contribution from
the bulge (\textit{e.g.} \cite{Boehm:2003bt,Vincent:2012an, review}).

This paper discusses a novel possibility. Positrons that escape from
their source and their host galaxy in jets or winds may reach the
intergalactic medium (IGM) to stay there indefinitely, as long as
their cooling time remains longer than the age of the universe.
Positrons of very high energy ($E \gtrsim 100 (1+z)^{-5/2} \,
{\rm MeV}$ at redshift $z$) are slowed by Compton
scattering with the CMB, and positrons of very low energies (highly
subrelativistic) cool by Coulomb scattering with thermal
electrons. At intermediate energies, however, positrons do not cool over
the age of the universe in the IGM. The turbulent magnetic structure of the escaping 
jets and winds can furthermore trap the pairs, forcing them to follow the motion of the plasma fluid as long as the magnetic
field has enough small-scale structure to prevent particles from
diffusing over large distances by moving along the field lines.
Positrons fill the IGM, and if they are eventually accreted by gravity
wells such as our own galaxy, they annihilate whenever they reach a
medium with high enough density to cool by Coulomb
scattering (e.g., \cite{oscar}).

 To the best of our knowledge, the intergalactic positron abundance
produced by astrophysical sources has not been previously considered. 
In this letter, we estimate the intergalactic positron density and show
that positrons accreted by the Milky Way may substantially
contribute to the Galactic 511 keV emission line. We first estimate the
IGM positron-to-electron ratio based on cosmic radio background (CRB)
observations, assuming that the radio signal is produced in sources
similar to the observed active galactic nuclei (AGN) jets and radio
lobes. We then compute the expected 511 keV line
luminosity and the emission profile in a spherical accretion model.
We find that the annihilation in this spherical model happens mostly in 
the galactic bulge as required by observations. We conclude with a
discussion of the expected changes of the emission profile in a more
realistic accretion model as well as the potential of new radio observations  
of the small scale fluctuations to test our hypothesis.

The CRB has been measured from 0.01 to 90 GHz
by several collaborations including, most recently, the ARCADE-2
experiment \cite{arcade2}. The observed brightness temperature spectrum,
in excess of the constant cosmic microwave background (CMB) and corrected for Galactic 
emission, is well fitted by a power-law:
 \begin{equation}
T_{\rm{radio}} = (1.26 \pm 0.09 \rm{K})
 \left( \frac{\nu}{\rm{GHz}} \right)^{-2.6 \pm 0.04}.
\label{arcade}
\end{equation}
The shape of the spectrum is reproduced by synchrotron emission
\cite{Gervasi,Cline:2012hb} from a population of electrons and positrons
with a power-law distribution of energies $E=\gamma m_e c^2$,
$n_p(\gamma)\, d\gamma \propto \gamma^{-p}\, d\gamma $,
with an index $p = 2.2$. 
Extrapolations from luminosity functions of
known synchrotron-emitting sources account only for about one sixth of
the observed background intensity \cite{Gervasi}. 
Underestimating the level of Galactic emission is a potential contaminant  
 \cite{Keshet:2004dr}. However, the expected contribution is determined with 
 tight errors, 5 (0.4) mK at 3.3 (10) GHz, compared to the CRB brightness 
 temperature of 54 (3) mK at the same frequencies.

 The brightness temperature in equation (\ref{arcade}) corresponds to
an energy density per unit frequency, $u_{\nu}$, of
\begin{eqnarray}
 \nu\, u_\nu &=& {8 \pi \nu^3 k_B \over c^3} T_{\rm radio} \nonumber \\
&=& (1.0\pm 0.1) \times 10^{-7}
 \left(\frac{\nu}{1 \mathrm{\, GHz}}\right)^{0.4} \rm{eV \ cm}^{-3}\, .
\label{rho}
\end{eqnarray}
Pairs with a Lorentz factor $\gamma$ radiating synchrotron in a magnetic
field $B$, with Larmor frequency $\omega_B=eB/(m_e c)$, dominate the
emitted power at an emitted frequency $\nu_e = A\gamma^2 \omega_B$ and
have a synchrotron cooling time $t_c=9c/(4\gamma r_e \omega_B^2)$. Here,
$r_e$ is the classical electron radius and $A$  is a constant close to
unity, which we use to match the integrated energy density in pairs
to the photon luminosity at a given frequency (see equation 4 below).
We assume an approximately constant magnetic field, and we define $t_e$ as the
time over which pairs in a radio source radiate via
synchrotron emission. This roughly corresponds to the
lifetime of the source. During this time, pairs with
Lorentz factor $\gamma$ will have radiated a fraction $t_e/t_c$ of their
rest mass energy. This may be used to relate the average number density
of radiating pairs with energy $\gamma m_e c^2$ at a mean source
redshift $(1+z_r)$ to the present
comoving energy density of emitted synchrotron photons:
\begin{equation}
\label{npf}
 n_p(\gamma) d\gamma = {u_{\nu} d\nu (1+z_r) \over 2\gamma m_e c^2}\,
 {t_c\over t_e}\, f_I \, ,
\end{equation}
where $f_I$ is the fraction of contributing pairs that were able to
escape to the IGM, and the observed
frequency is $\nu=\nu_e/(1+z_r)$. Note that $t_e/t_c$ is much less than unity for low-$\gamma$ pairs that radiate only a small fraction of their energy
to the radio background, and can be larger than one for very high
$\gamma$, when pairs need to be reaccelerated many times within a
source in order to maintain the emitted power-law spectrum. Now,
using $u_\nu\propto \nu^{-(p-1)/2}$ and
$\gamma=[\nu(1+z_r)/(A\omega_B)]^{1/2}$, and computing the total number
density of pairs $N_p= n_p(\gamma)\gamma^p/[(p-1) \gamma_{min}^{p-1}]$
integrated for all $\gamma > \gamma_{min}$, we infer from
equation (\ref{npf}) that
\begin{equation}
\label{nps}
 N_p = {\nu u_\nu \over m_e c^2} 
\left({A \omega_B\over \nu}\right)^{3-p\over 2}
{9c\, f_I (1+z_r)^{p-1\over 2}\over 4(p-1)r_e \omega_B^2 t_e
 \gamma_{min}^{p-1}} ~.
\end{equation}
This relationship tells us that the number density of pairs
is equal to the energy density per unit $\log\nu$ in the radio
background divided by the electron rest-mass energy, times a number of
dimensionless factors. Note that $N_p$ is independent of the frequency
$\nu$ at which we choose to evaluate it, because $\nu u_\nu\propto
\nu^{(3-p)/2}$. The uncertainties are the magnetic field, the emission
time $t_e$, the mean source redshift $z_r$ and the escape fraction $f_I$. The factor $\gamma_{min}$ should be
close to unity in the source rest-frame for a realistic mechanism to
accelerate the synchrotron emitting particles.

  Using typical values for the lobes of radio-loud AGN of $B\sim
10\,\mu{\rm G}$ and $t_e\sim 10^{7.5}\, {\rm yr}$
(assuming that radio lobes expand at the typical intracluster sound speed),
replacing $p=2.2$ and using $z_r=1$ and $A=1$,
and dividing by the mean electron density in the universe at present,
$\bar n_e=2.3\times 10^{-7} \, {\rm cm}^{-3}$ \cite{PDG},
we obtain the IGM positron-to-electron ratio: 
\begin{equation}
\label{npt}
 N_p/\bar n_e = 1.6\times 10^{-5}
\left( {B \over 10\, \mu{\rm G}} \right)^{-1.6}
{10^{7.5}\, {\rm yr} \over t_e} {f_I\over \gamma_{min}^{1.2}} ~.
\end{equation}
Thus, for typical parameters of a radio lobe,
$N_p/\bar n_e\simeq 10^{-5} f_I$.
For the core regions of AGN jets there is an additional Doppler factor
of $\delta=[\Gamma(1-\beta\cos\theta)]^{-1}$ in the observed luminosity
due to relativistic beaming, where $\Gamma$ is the jet bulk
Lorentz factor and $\theta$ is the angle between the jet and the line of
sight. Equation (\ref{npt}) is then modified with 
a factor $\delta^{-\frac{5-p}{2}}$. 
The typical value of the magnetic field in the core of AGN jets is $B\sim 3\, {\rm mG}$, as estimated from a sample 
of resolved sources \cite{ghise93} showing a characteristic spectrum of synchrotron 
radiation and self-absorption \cite{marsch83}. For the same sources $\delta \sim 8.5$ and the emission time is the light-crossing time 
of the core region, $t_e\sim 25\, {\rm yr}$. We obtain the IGM
positron-to-electron ratio $N_p/\bar n_e \sim 10^{-5}\, f_I $. Note that
many radio jets may be produced in isolated galaxies without a massive
intracluster medium surrounding them, so the relativistic jet may be
directly expelled to the IGM without producing observable radio lobes.

We conclude here that if most of the CRB were generated by
electron-positron pairs in sources similar to the AGN jets and radio
lobes (with similar values of $B^{1.6}t_e$), and if most of the
positrons contained in these sources eventually escaped to the IGM, then
there would be about ten positrons for every million electrons in the
universe. This fraction may be much lower if the radio background is
mostly produced by electrons rather than pairs, or from sources with
high values of $B^{1.6}t_e$ compared to AGN, or if most of the positrons
do not escape. We note that these energetic particles would contribute to the matter
pressure of the IGM. The ratio of the energetic particles pressure to
the thermal IGM pressure is $\sim N_p m_ec^2/ (\bar n_e k_B T_I)\sim 5$,
for $T_I=10^4$ K and $N_p/\bar n_e\sim 10^{-5}$, so for the maximum
value of the positron abundance the pressure would appreciably modify
the IGM dynamics determining the properties of the Ly$\alpha$ forest.

Such a large $e^+$ fraction could lead to a visible annihilation signal as these positrons are accreted 
into galaxies with other intergalactic matter. In fact, if our galaxy is accreting matter at a rate $\dot M$, the
number of positron annihilations implied is $10^{49.5}\, (N_p/\bar n_e)\,
\dot M/(M_\odot\, {\rm yr}^{-1})\, {\rm s}^{-1}$. For the Milky Way,
we can expect a gas accretion rate of $\sim 1\,M_{\odot}\,{\rm yr}^{-1}$ \cite{Kennicutt:2012ea},
and reproducing the observed INTEGRAL 511 keV luminosity implying an
annihilation rate of $2\times 10^{43}\, {\rm s}^{-1}$ therefore requires
$N_p/\bar n_e\sim 10^{-6}$, which is safely below our estimated
maximum and can be matched by assuming $f_I\sim 10^{-1}$.

  Next, we model the distribution of the 511 keV emission from spherically
symmetric accretion of plasma containing relativistic positrons. The
dominant contribution to the positron cooling rate is due to Coulomb
scattering, and is roughly proportional to the density of electrons in
the interstellar medium (at high positron energies, the cooling rate does
not depend much on whether the electrons are in ionized, atomic or
molecular matter). A positron with an initial energy $E_{\rm  IGM}$
before infall will annihilate when it reaches a region of high enough
density for it to efficiently thermalize; this location depends on the
matter distribution and the energy $E_{\rm IGM}$.

  We parameterize the matter distribution in the Milky Way as an axially
symmetric distribution of matter, using a model that compiles the results
of several observation studies described in detail in
Ref.\,\cite{gasDistr}. The height-dependence of the H\textsc{i} and
H$_2$ densities is
Gaussian, with scale heights of 250\,pc and 70\,pc, respectively. The
H\textsc{i} density on the galactic plane at our location is $\sim 1$
atom cm$^{-3}$, whereas H$_2$ is in a molecular ring around a
galactocentric radius of $R = 5$\,kpc, with a peak density on the plane
of $\sim 2.5$\,atom cm$^{-3}$. The H\textsc{ii} distribution has
two components, a central exponential disk with a scale height of 1\,kpc
and central density of 0.025\,atom cm$^{-3}$, and an annulus around
$R = 2$\,kpc representing H\textsc{ii} regions, with a scale height of
150\,pc and peak density of 0.2\,atom cm$^{-3}$. We use the
energy loss-rates computed with the GALPROP package
\cite{galprop_url,Strong:2007nh}, which also includes the subdominant
effects of inverse-Compton scattering of CMB and interstellar light,
bremsstrahlung, synchrotron radiation and ionization. Details of the
exponential magnetic field model and the interstellar radiation field
used to compute these are also in Ref.\,\cite{gasDistr} and references
therein. 

  The positrons are initially distributed in a power law with index
of $-2.2$, following the inferred synchrotron-producing spectrum from
the CRB, with lowest energy of 1 MeV. We have checked that decreasing 
the index to $-2.5$ (as in spectra observed in most AGNs) has little quantitative
impact. Note also that positrons of very high energy should have
cooled below $\sim 10$ MeV in the IGM by Compton scattering, although
these are a minority of the particles and it also has little impact on
the computed annihilation distribution.
The calculation of energy loss is stopped when positrons reach energies
below 100 eV, at which point thermalization and annihilation should
quickly follow. The flux of 511 keV photons can then be found from the
annihilation rate per unit volume $dN_{e^+}/dVdt$:
\begin{equation}
\ud \Phi_{511} = 2 \frac{\ud \Omega}{4\pi}\int_{\rm l.o.s.} \frac{1}{4} \frac{dN_{e^+}}{dVdt}(x) \ud x,
\end{equation}
where $x$ is the direction along the line of sight (l.o.s.), the prefactor of 2 accounts for the two outgoing photons per annihilation and the $1/4$ accounts for the fact positronium decay to two photons only occurs when para-positronium is formed, one quarter of the time.

\begin{figure}[h]
\begin{minipage}{0.45\textwidth}
\includegraphics[width=1.1\textwidth]{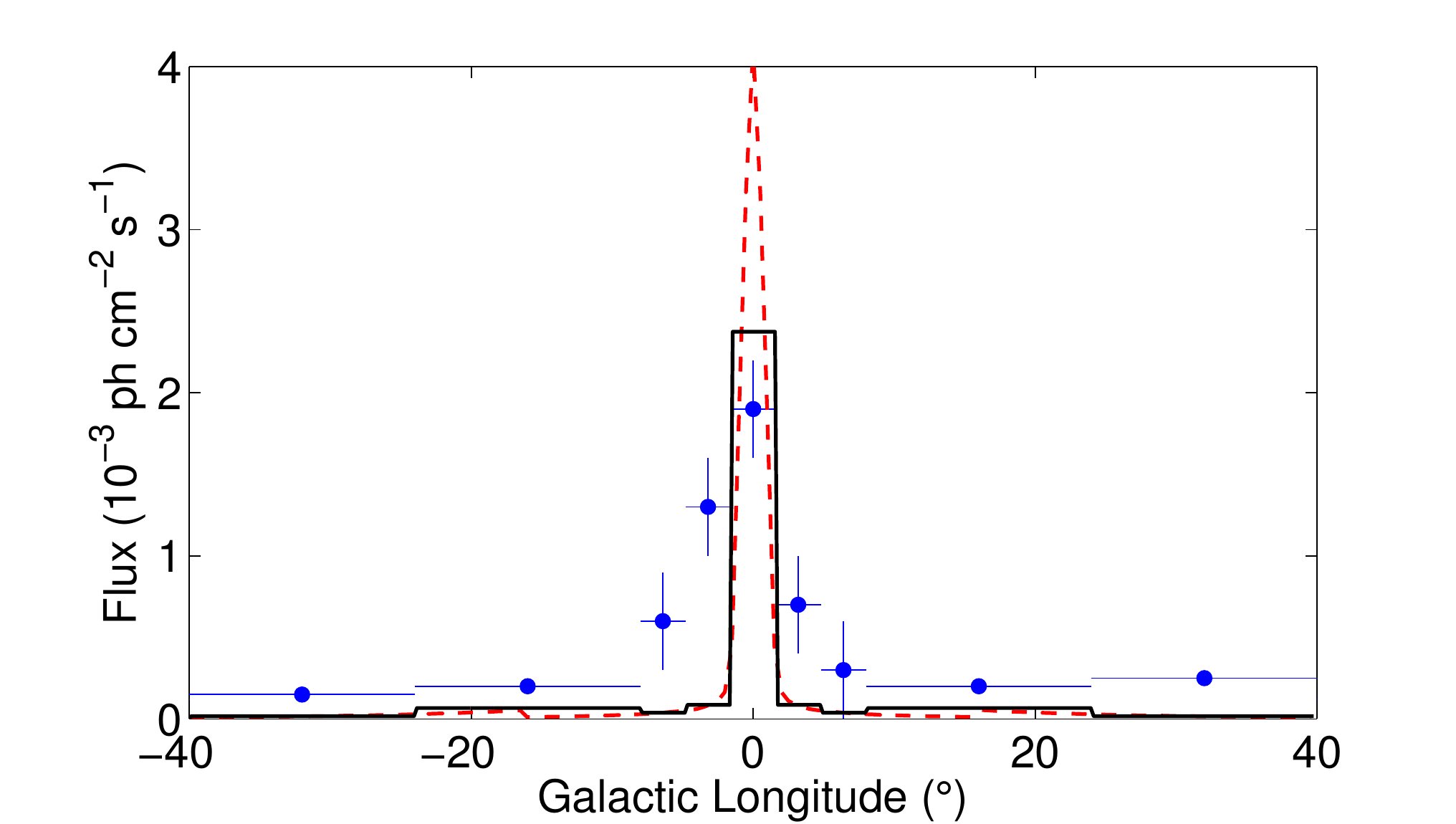}
\includegraphics[width=1.1\textwidth]{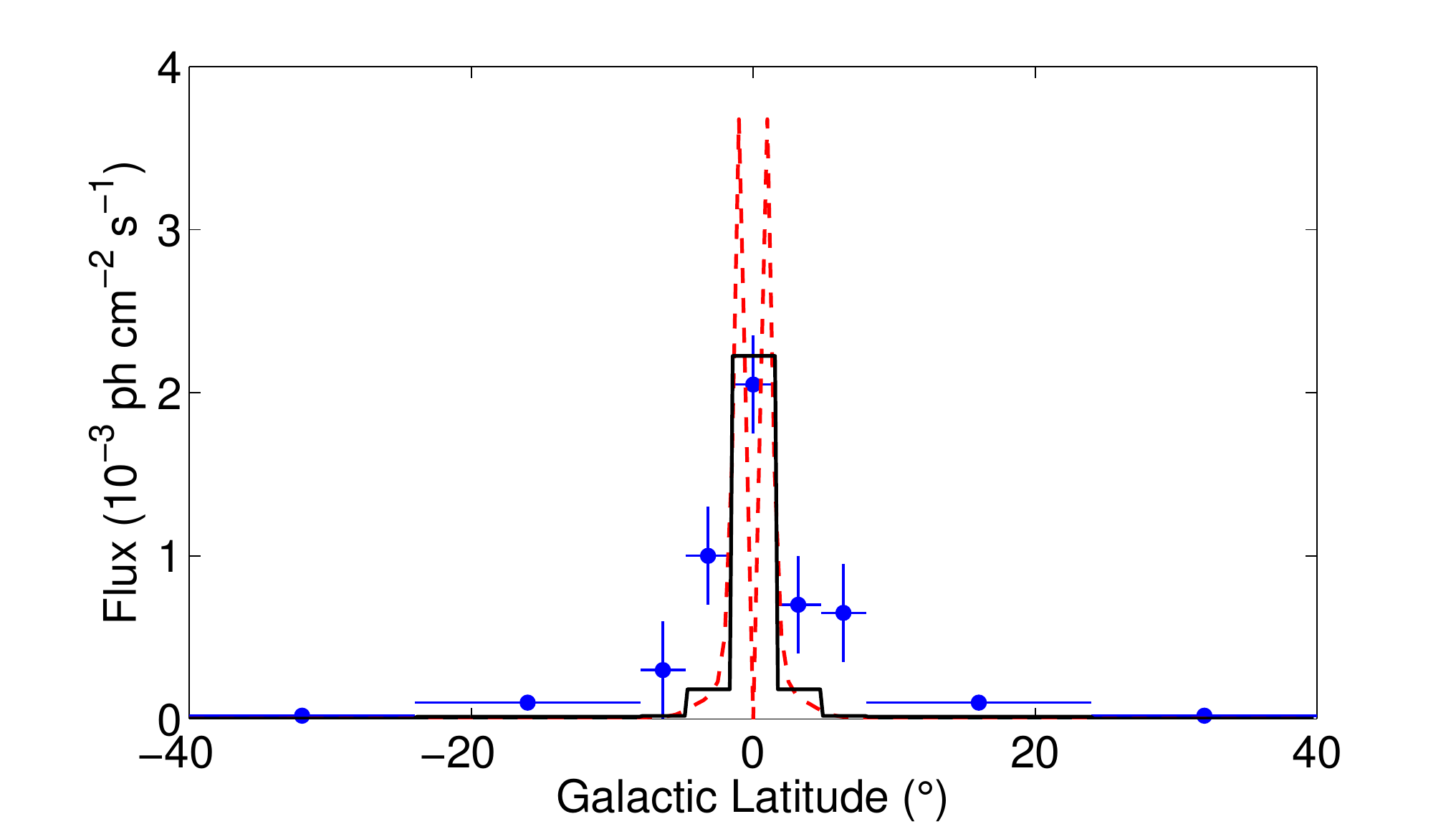}
\end{minipage}
\caption{\textit{The binned INTEGRAL/SPI 511 keV data from \cite{integral},
integrated over $|${\rm Latitude}$| < 8^\circ$ $(|${\rm Longitude}$|$$< 8^\circ)$,
is shown by the data points in blue in the upper (lower) panel. The
outer four bins are $16^\circ$ wide, and inner bins are $3.2^\circ$
wide. Black solid line shows the predicted flux in the spherical infall model,
binned in the same way. The red dashed line shows the shape of the
predicted flux, with its area matching the normalization of the histogram.}}
\label{integral2}
\end{figure}

The results  are then compared with the INTEGRAL/SPI observations.
Figure \ref{integral2} shows a comparison between measured and predicted
flux in the inner 16 degrees of latitude (upper panel) and longitude
(lower panel). These figures illustrate that the spherical accretion
model allows us to reproduce the large annihilation signal from the
bulge. Using the parametrization of Ref.\,\cite{Weidenspointner2008},
we obtain a bulge-to-disk ratio of 2.6, leaving room for annihilations from
Al$^{26}$ produced in supernovae and other $\beta^+$-producing isotopes
distributed in the disk.

Spherical accretion concentrates the infalling positrons close to the
Galactic centre, producing a bright annihilation region in the bulge. A more
realistic model would allow for the IGM gas to carry angular momentum
and for a non-spherical galactic potential, so the gas could then fall
over a much larger area around the disk.
This would redistribute the luminosity over the disk, reducing the central luminosity and smearing 
the small-scale features. The positrons annihilate at the point where
the medium is dense enough for them to lose sufficient energy, and this
may often occur in disk crossings rather than near the bulge. However,
the details depend on the way the structure of the magnetic field in
the Galaxy halo may modify the trajectories of accreting IGM plasma.
Previous studies have considered a possible focusing effect of a dipole magnetic field 
lines of the halo, although there has not been direct observation of such a dipole 
component \cite{Prantzos:2005pz,review}. 
Nevertheless, if such a dipole field is present, it could contribute to an enlarged 
positron density in the bulge.
This is clearly a complex problem that we cannot address in this
letter. We note also that if the extragalactic positrons contribute
substantially to the total annihilation signal but in a more extended
fashion, the concentration to
the bulge might also be caused by recent AGN activity from the central
black hole in the Milky Way having delivered positrons to the bulge region.

The sources of the CRB are still unknown. It has recently been argued \cite{Holder:2012nm} that the RMS fluctuations of the CRB are more than an order of magnitude smaller than those of the infrared background. If cosmological, the sources for the radio background might be either at high redshift or spatially extended (on Mpc scales), 
so that  the clustering amplitude is substantially lower. Alternatively,
the radio sources might not be highly biased, for example if they arise
from a population of low-luminosity AGN in low-mass halos, and this would
reduce their clustering.
The new generation of radio observations, measuring the small scale fluctuations of the background, would be able to test whether the cosmological 
population of AGN cores and lobes are sourcing the 
radio background.  We note that there might be positrons injected in the IGM at redshifts higher than $\sim 6$, which would 
 Compton-cool to non-relativistic energies in the IGM before they are accreted
to a galaxy, so that at present they would annihilate at lower gas
densities leading to a 511 keV emission 
extending to higher galactic latitude than observed by INTEGRAL. 

In summary, we have presented a new possible source of positrons that
can contribute to the 511 keV emission from the Milky Way: the accretion
of intergalactic material containing positrons that have been produced
over the age of the universe in high-luminosity objects that can expel
matter to the IGM, such as the jets of active galactic nuclei. Based on
the radio background intensity, we have obtained their $e^\pm$
contribution to
the IGM using Eqs.\,(\ref{nps}) and (\ref{npt}). If the CRB comes from these synchrotron-emitting sources, we estimate a
maximum density of $10^{-5}$ positrons per electron in the IGM. The positron fraction required to
produce the observed INTEGRAL 511 keV luminosity is $\sim 10^{-6}$,
well below our maximum estimation. The main feature of the INTEGRAL/SPI
observation, namely the large density of positrons in the Galactic
bulge, can be reproduced by a simple spherical model of accretion,
although a prediction of the exact line emission
morphology will require more realistic modelling of IGM gas accretion and of the magnetic field structure in the
Galactic halo.

\acknowledgments 
 We thank R. Lineros, P. Martin, A. Marscher, N. Prantzos, A. Shalchi, E. Waxman and W. Xue for valuable discussions. 
 C.P-G is   supported  by  the   Spanish  MICINN   grants  FPA-2007-60323,
FPA2011-29678, the Generalitat Valenciana  grant PROMETEO/2009/116 and the ITN
INVISIBLES (Marie Curie Actions, PITN-GA-2011-289442). A.V. acknowledges support from the European Social Fund-European
Commission and from Regione Calabria. A.C.V. acknowledges support from FRQNT and European contract FP7-PEOPLE-2011-ITN and INVISIBLES.
J.M. is supported in part by Spanish MICINN grant AYA2012-33938.

\bibliography{./positrons.bib}

\end{document}